\documentclass[paper,twocolumn,twoside]{geophysics}

\usepackage{amsmath,amssymb,amsfonts,graphicx,subfigure,bm,yfonts}
\usepackage[hidelinks]{hyperref}
\usepackage{cleveref,xcolor}
\usepackage{rotating}
\usepackage{multirow}

\def\rmrkok#1{}

\def\figref#1{Figure~\ref{fig:#1}}
\def\figrefp#1{(Figure~\ref{fig:#1})}

\def\beq{\begin{eqnarray}}
\def\eeq{\end{eqnarray}}

\def\phi{\varphi}

\usepackage{amsmath}

\newcommand{\afl}[1]{\footnotemark[{#1}]}
\graphicspath{{./Fig/}}
\begin{document}

\title{Deep learning enhanced initial model prediction in elastic FWI: application to marine streamer data}
\author{Pavel Plotnitskii\afl{1}, Oleg Ovcharenko\afl{1}, Vladimir Kazei\afl{2}, Daniel Peter\afl{1}, Tariq~Alkhalifah\afl{1} \\
\afl{1} King Abdullah University of Science and Technology, Thuwal, Saudi Arabia \\
\afl{2} Aramco Americas, Houston, USA}
\lefthead{Plotnitskii et al.}
\righthead{\emph{Deep learning enhanced initial model prediction}}
\maketitle

\begin{abstract}
Low-frequency data are essential to constrain the low-wavenumber model components in seismic full-waveform inversion (FWI). However, due to acquisition limitations and ambient noise it is often unavailable. Deep learning (DL) can learn to map from high frequency model updates of elastic FWI to a low-wavenumber model update, producing an initial model estimation as if it was available from low-frequency data. 
We train a FusionNET-based convolutional neural network (CNN) on a synthetic dataset to produce an initial low-wavenumber model from a set of model updates produced by FWI on the data with missing low frequencies.
We validate this DL-fused approach using a synthetic benchmark with data generated in an unrelated model to the training dataset. Finally, applying our trained network to estimate an initial low-wavenumber model based on field data, we see that elastic FWI starting from such a “DL-fused" model update shows improved convergence on real-world marine streamer data.
\end{abstract}

\section{Introduction}
Full-waveform inversion (FWI) seeks to match synthetic and observed seismic waveform data through iterative, incremental refinements of a subsurface model, minimizing a pre-defined data misfit functional \citep[e.g.][]{virieux2009}. Being able to handle media of arbitrary complexity, FWI applications dominate modern seismic imaging in complex media, in particular for marine environments \citep[e.g.][]{warner2013, shen2018,skopintseva2019regularization}.

One key drawback of FWI is the need for low frequencies \citep{baeten2013} to deliver reliable results. This is due to mainly three causes: (i) the employment of gradient optimization schemes in FWI, (ii) non-convex data misfit functionals with multiple local minima, and (iii) the lack of accurate initial subsurface estimates. Different modifications have therefore been proposed to improve full-waveform inversion (FWI) workflows without the need of low-frequency data and corresponding initial low-wavenumber information.   

A prevailing approach is to modify the data misfit functionals \citep[e.g.][]{luo1991wave,choi2012,leeuwen2013,sun2019robust,fuqiang_doublediff_2022}, which can significantly improve the convergence of FWI.
For example, optimal transport misfits \citep{metivier_2016,fuqiang_2018,fuqiang_2018_entropy}, differentiable dynamic time warping misfits \citep{fuqiang2021MisfitFB,fuqiang2022}, or traveltime-based misfits \citep{schuster_traveltime} might overcome cycle-skipping problems in FWI.
However, a fundamental lack of sensitivity in seismic data has to be populated with additional assumptions and constraints.

Another approach to address the ill-posedness of the inverse problem is to apply model regularizations. For example, filtering model updates allows for a more direct incorporation of prior assumptions about low-wavenumber updates into FWI \citep{ravaut2004multiscale}, potentially resembling those properties of FWI updates at low frequencies \citep[e.g.][]{alkhalifahFullmodelWavenumberInversion2016, kazei2016, kalita2018}.
Regularization techniques, such as total variation (TV) or Sobolev space norms \citep[e.g.][]{esserTotalvariationRegularizationStrategies2016, kazeiSaltbodyInversionMinimum2017, kalita2019regularized, skopintseva2019regularization} allow to constrain low-wavenumber model updates in FWI based on prior model assumptions when inverting the data. Similarly, diffusion filters and structure-preserving shaping approaches have been proposed to address model smoothing that incorporates geological structure information \citep{Fehmers2003,Fomel2007,Lewis2014,kazei2016,Trinh2017,Yao2019}. These approaches typically rely on prior structural assumptions or single model information. In the following, we will explore the possibility to use image information coming from multiple FWI gradients to estimate a low-wavenumber model update for realistic datasets.

Recently, deep learning (DL) approaches have shown promising progress in seismic data preprocessing 
%
%
\citep[e.g.][]{ravasi2021joint,ovcharenko_multiples_separation_2021,birnie2021_2_geoph,alali_TL_2021}, 
as well as in seismic inversions \citep[e.g.][]{zhang_2019_reg_fwi,cnn_boost_paper,aragao2020elastic,sun2020mlmisfit,li2021_dl_wl_efwi,kazei2021predicting}. 
Conventional FWI algorithms can be coded as a physics-based deep neural network training procedure \citep{richardson2018seismic}.
DL thus offers a variety of solutions to different challenges in FWI, from completely replacing FWI with supervised DL to modifying either data inputs or model outputs and deeply entangling workflows of FWI and DL.

Focusing on low-wavenumber information, it can be derived directly from seismic data in the form of low-frequency data extrapolations \citep{Demanet_li_2016,ovcharenkoNeuralNetworkBased2017, ovcharenkoLowFrequencyDataExtrapolation2018, jin2018learn, sunLowFrequencyExtrapolation2018,hu2019progressive,sun2020extrapolated, sun2021deep} using deep learning tools. \cite{ovcharenko2021transferring, ovcharenko_lowfreq_land_dualband_2021} showed that low-frequency seismic data extrapolation can be approached in a multi-task learning framework and conducted experiments on real marine streamer data, bridging the gap from synthetic to real marine and land data.

Alternatively, deep learning approaches can infer velocity models directly from seismic data \citep[e.g.][]{polo2018,yang2019deep,oye2019velocity,kazei_2020_elastic_near_surface_model_estimation}. While results for low-dimensional synthetic models are comparable to conventional FWI \citep{farris2018tomography, araya2020fast, kazei2021predicting}, subsurface models still need to be similar to the models from the training data sets. This highlights a major problem in supervised DL approaches in that the selection and generation of an appropriate training set becomes crucial for the inference on real data.

Here, we explore a deep learning approach in the image domain to predict a low-wavenumber model update based on multiple FWI model updates from the seismic data that lack low frequencies. In \cite{plotnitskii_aegc_2019}, the authors utilized a convolutional neural network (CNN) and a one-dimensional trace-by-trace approach for the prediction of low-wavenumber models. The authors were modeling low- and high-wavenumber models with a wavelet synchro-squeezed transform (WSST) filter and seismic wave propagation theory. In a subsequent step, \cite{plotnitskii2020_eage} mapped high-wavenumber updates to low-wavenumber ones for a synthetic acoustic example.
Here we advance this concept to elastic FWI updates and field data from marine streamers. 

Our goal is to further develop and apply the low wavenumber prediction to real data.
First, we explain the proposed method for recovering the low-wavenumber structure of the subsurface by deep learning. Then, we introduce a real marine streamer dataset and our approach for generating synthetic data. After that, we explain the frameworks for training a deep learning model and running numerical experiments. Finally, we apply the method for elastic FWI initialization on synthetic and real-world datasets.

\section{Method overview}
We directly map high-wavenumber model updates from a time-domain FWI into a suitable low-wavenumber model update using a deep neural network, as shown in \figref{method}.
The estimated low-wavenumber model update is then added to the initial model to construct a new, predicted initial model. The proposed DL method operates on by-products of FWI and, thus, might be incorporated into existing workflows in a straightforward way.
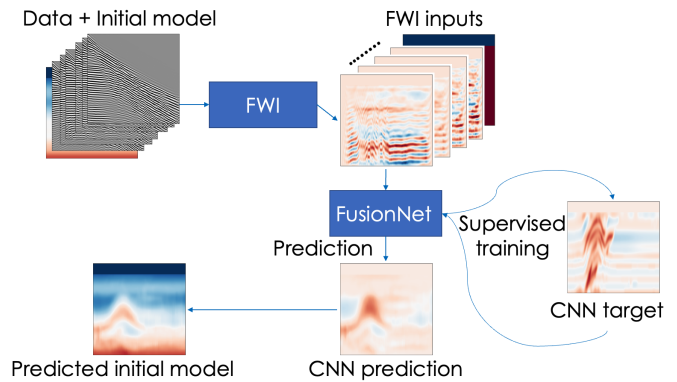
\plot{method}{width=\linewidth}{Workflow for low-wavenumber extrapolation by a deep learning model. A set of 10 high-frequency FWI updates, initial model and water taper are mapped into the target low-wavenumber update.}
In deep CNNs, convolutional kernels in shallow layers are capturing high-frequency patterns in the input data while the cumulative receptive field constructed in deeper convolutional layers describes longer wavelengths in the input data. Here, we assume an inaccurate initial model for inversion that leads to globally displaced reflectors in migrated images. Thus, we rely on global patterns captured by a FusionNet convolutional architecture \citep{fusionNet} to recover smooth velocity variations at lower wavenumbers.


\section{Datasets}
For supervised machine learning applications, it is vital to construct realistic training and validation datasets. Specifically, the training dataset should be sufficient to enable knowledge transfer between training (synthetic dataset) and application domains (field dataset). Thus, features and distribution of samples in the synthetic dataset should contain respective quantities in the field data. We, therefore, use the survey design from the field dataset to generate synthetic data.

\subsection{Marine streamer data}

The 2D marine streamer dataset used to test our method was acquired in the Northwestern part of the Australian continental shelf by CGG, employing the BroadSeis acquisition system with a variable depth streamer \citep{soubaras2010variable,Soubaras2012BroadSeisE}. We generally follow the approach described in \cite{kalita2017efficient} to process this dataset.

There are 1,824 common-shot gathers in the survey with an approximate source spacing of 18.75~m. The streamer contains 648 hydrophones with an approximate group interval of 12.5~m. Thus, the near and far offsets are about 169 and 8300~m, respectively. Time sampling of the data is 2~ms, and the total length of recording is 6.2~sec. Because of the poor signal-to-noise ratio, the dataset is preprocessed by CGG such that frequencies below 2.5~Hz are filtered out. \figref{data_field_synth}a shows a single common-shot gather from this dataset, with its average power spectrum displayed in \figref{data_field_synth}b. \figref{data_field_synth}c to \figref{data_field_synth}f show a corresponding synthetic common-shot gather from a training model. The source wavelet for each source location is estimated in the frequency domain following \cite{kim2011frequency}. For the sake of simplicity, we assume the average wavelet signature to be shared among all sources, and chose the first 576 shot gathers to reduce the size of the dataset. 

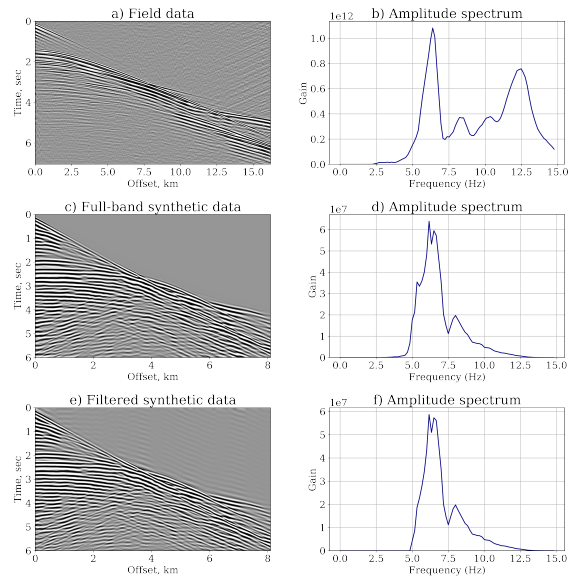
\plot{data_field_synth}{width=\linewidth}{Field data: a) real shot gather, b) real shot gather spectrum.
Filtered synthetic shot gather: c) full-band shot gather, d) full-band shot gather spectrum, e) high-pass filtered shot gather, f) high-pass filtered shot gather spectrum.
}

\subsection{Synthetic data}

Our data generation routine builds multiple input-target pairs for supervised training. We construct the input data, high-wavenumber FWI model updates, by running inversions of synthetic seismic data which were band-passed within the 5-10~Hz range. The target data, low-wavenumber model updates, are built in two steps. First, we remove high wavenumbers from the true velocity model by smoothing it with a Gaussian filter. Then, we build the final target data by subtracting the initial model from the smoothed true model.

\subsubsection{Random subsurface models}

There is no trivial approach to generate realistic geological models for supervised training. For example, previous studies utilized deep neural networks \citep{ovcharenko2019style,wu2020building} or wavelet packets \citep{kazei2019realistically} for the generation of realistic geologic models. 
In this work, we rely on an approach which combines the creation of random layered velocity models with constrained trends and elastic image transformations \citep{simard2003best}. Our velocity model generator produces folded layered models by elastic transforms of 1D velocity models. It is modified from \cite{ovcharenko2021transferring} to fit velocity boundaries assumed in our seismic dataset. This simplifies the application of the method where only one or few well logs are available. However, more complex models could be generated in a similar fashion to deliver higher-fidelity datasets \citep{kazei2021predicting}.

First, we create a 1D linear Vp velocity model with random gradient between 0.33 and 0.8~km/s per km to cover the range of trends in our velocity models. Second, we create a 1D sequence of reflection coefficients (velocity perturbations), then transform them to P-wave velocities.
Third, we add the perturbations to the 1D linear velocity model and then apply a distortion map from a random Gaussian field to modify the velocity model.
Finally, we apply lower- and upper-velocity constraints to the modified velocity model. \figref{random_v} shows four examples of such generated velocity models. As a result, each model contains a water layer of random thickness and relatively flat layered geology, which is very common for sedimentary environments. The velocity generator, which we use, provides water taper (the precise location of water in velocity model) alongside with produced water model. In the case of a real field dataset in industry, exploring the water bottom topography does not cause any significant problems. This is the reason, why we assume the known geometry of the sea floor.

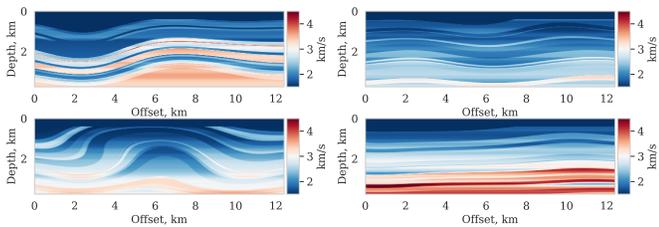
\plot{random_v}{width=1\columnwidth}{Examples of synthetic  $V_p$ velocity models from the training dataset.}

For our elastic FWI application, we produced a dataset comprising 3,904 models of size $496 \times 150$, which corresponds to a model dimension of $12,400 \times 3,750$ meters with a grid spacing of 25~m. The generator outputs both a "true" velocity model and a 1D velocity trend from it that can be used as initial model for FWI. To construct density, $\rho$, and shear velocity, $V_{s}$, from the generated $V_{p}$ model values, we use Gardner's relation \citep{gardner_relationship}
\begin{equation}
    \rho = \alpha \, V_{p}^{\beta}
\end{equation}
with $\alpha=0.3$, $\beta=0.25$ and a ratio $V_{p}$/$V_{s}=\sqrt{3}$ for Poisson solids.
In \figref{vel_models}, we show that the set of synthetic models evenly spans the range of benchmark velocity models. Specifically, we plot well-logs extracted from the mean offset of the field dataset as well as the Marmousi and Overthrust benchmark velocity models. These are then overlaid with a set of randomly selected logs from synthetic velocity models in the training dataset.

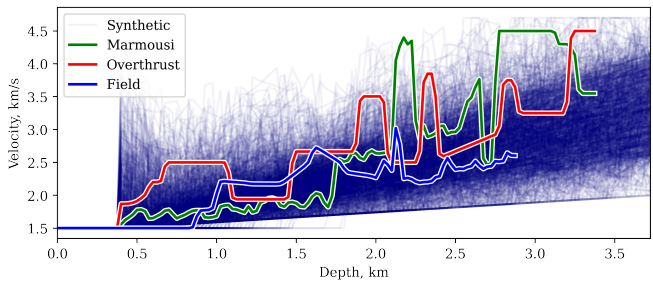
\plot{vel_models}{width=1\columnwidth}{Well-logs extracted from the field dataset (blue), the Marmousi (green) and Overthrust (red) benchmark velocity models, overlayed with randomly selected logs from generated $V_{p}$ velocity models.}

\subsubsection{Wave propagation}

At the next stage, we generate synthetic waveform data using the following parameters extracted from the field data:
\begin{itemize}
    \item Source wavelet from a real marine survey.
    \item Shot gather length 6~s, time sampling dt=2~ms.
    \item Grid spacing=25~m, source spacing=200~m, receiver spacing=25~m.
\end{itemize}
For this exploration setup, we treat the data below 5~Hz as the sought for low frequencies. To simulate the lack of low frequencies in our synthetic seismic dataset, we apply the high-pass filters with the corner frequency of 5~Hz to the set of generated synthetic shot gathers. \figref{data_field_synth}c and \figref{data_field_synth}e show the full-band and band-passed shot gathers from the Marmousi velocity model, respectively. \figref{data_field_synth}d and \figref{data_field_synth}f show the spectra of corresponding shots.
We use an open-source elastic FWI code \citep{kohn2012,kohn_phdthesis} to run finite-difference wave propagation and inversion in the time domain with a fourth-order spatial and second-order temporal discretization.
FWI model updates are computed using an L-BFGS scheme with 20 memory steps.
The software \citep{kohn2012} allows to customize the inversion workflow by e.g. frequency filtering, spatial filtering of the gradients, time windowing, offset windowing, preconditioning, different misfit functions. These techniques aid to tackle the nonlinearity of the inversion process.
Also, it is possible to run inversions in a multi-stage manner meaning that the iterative process can be divided into stages with different parameters in each stage.

Recall, we need to generate high-wavenumber model updates to be the input for CNN training. To do that, we run the inversion for 10 iterations on the available seismic data with lack of low frequency data. This number of iterations appears to be sufficient to resolve the general features in the model, further increasing of number of iterations is not justified due to the large amount of required computing resources.
For generation of CNN inputs, we divide the inversion into 2 stages. The first stage contains 5 iterations with low-pass filter below 8 Hz applied. The second stage contains 5 iterations with low-pass filter below 10 Hz. By applying the low-pass filters sequentially with the increasing cut-off frequency we allow the inversion to start with narrow-band low-frequency data, which restores the most important low wavenumbers in first iterations. We set the $L_2$ norm as the objective function at all stages of inversion.

With respect to the computational costs of generating such a training dataset, all simulations ran on CPU cores, parallelized across shots and physical domain. With 160 computational cores and the acquisition parameters described above, it took approximately 20-40 minutes to finish 10~iterations of FWI for one model. Therefore, a significant amount of computational resources is needed to create the dataset consisting of ~4000-5000 velocity models within a few days.

The complete training dataset then consists of the generated random velocity models together with their 1D background trends, 10 consecutive $V_p$ model updates from the corresponding FWI iterations, and the smoothed version of the initial models as targets.

\section{Deep Learning Framework}

A deep neural network is known to be a universal approximator that might fit any non-linear relation in the data given a sufficient number of neurons \citep{Cybenko1989ApproximationBS}. Supervised deep learning relies on known inputs and outputs to learn the mapping between them by optimizing a specified kind of metric. Therefore, we begin by describing the design of the training dataset, then look into the selection of metrics used for training and trained model evaluation, and discuss the neural network architecture. Finally, we show the training and inference performance of the network.

\subsection{Setup}

To find a suitable training input, we performed an ablation study. Specifically, we tested different configurations of CNN inputs before choosing the preferred one. There are available initial velocity model (referred as "I", \figref{marm_all_channels_short}c), high-wavenumber model updates from 10 iterations of FWI (referred as "G", \figref{marm_all_channels_short}a-b), 
and the known water mask (referred as "W", \figref{marm_all_channels_short}d). In \figref{marm_all_channels_short}a-b, we show only the 1st iteration model update and 10th iteration model update as all 10 iterations of model updates look generally similar.
To construct different combinations of inputs with the high-wavenumber model updates, we thus have the possibilities: G (only high-wavenumber model updates), GI (high-wavenumber model updates \& a priori initial model), GW (high-wavenumber model updates \& water taper), and GIW (high-wavenumber model updates \& a priori initial model \& water taper). The combination with the best results on our DL tests was GW. 
Thus, in our deep learning framework, a set of high-wavenumber model updates and water taper serves as a single input sample for CNN training, whereas the low-wavenumber model update serves as a single target (\figref{marm_all_channels_short}e).

\subsubsection{Training dataset preparation}\label{data_for_dl}

Thus, we construct the 11-channel input data, by concatenating FWI model updates from the first 10 iterations ({\figref{marm_all_channels_short}a-b}) together with the water taper (\figref{marm_all_channels_short}d).
The network then aims to reconstruct the low-wavenumber target update, shown in  \figref{marm_all_channels_short}e.
In total, we created a dataset comprising 3,904 samples. We found it essential to take out outliers from the created dataset prior to training the network. In particular, there were generated velocity models which produced high-wavenumber model updates with very small amplitudes, in which case we removed those models from the training dataset. 

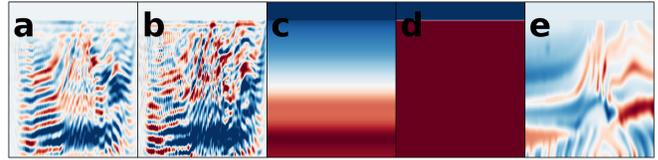
\plot{marm_all_channels_short}{width=\columnwidth}{Data available for DL training (12 input channels, 1 target channel): a)1st iteration model update (referred as G), b)10th iteration model update (referred as G), c)initial model (referred as I), d)water taper (referred as W), e)target low-wavenumber model update.}

A histogram of the curated dataset is shown in \figref{hists}. It represents the distribution of velocity perturbations in the generated dataset. \figref{hists}a and \figref{hists}b show the distribution of high-wavenumber model updates from the 10th iteration and target low-wavenumber model updates, respectively. 
\newline
The mean and standard deviation values for the high-wavenumber model updates (\figref{hists}a) and for the low-wavenumber model updates (\figref{hists}b) are equal to $-11,67$ and 28,241 m/s, correspondingly. It means that the distribution of high-wavenumber model updates is slightly negatively skewed and the distribution of low-wavenumber model updates is positively skewed.
Generally, we observe that despite of the described weak skewness of histograms the majority of values group around zero.
The last 11th channel of the CNN input is the water taper, it has values from 0 to 1 and it does not need scaling.
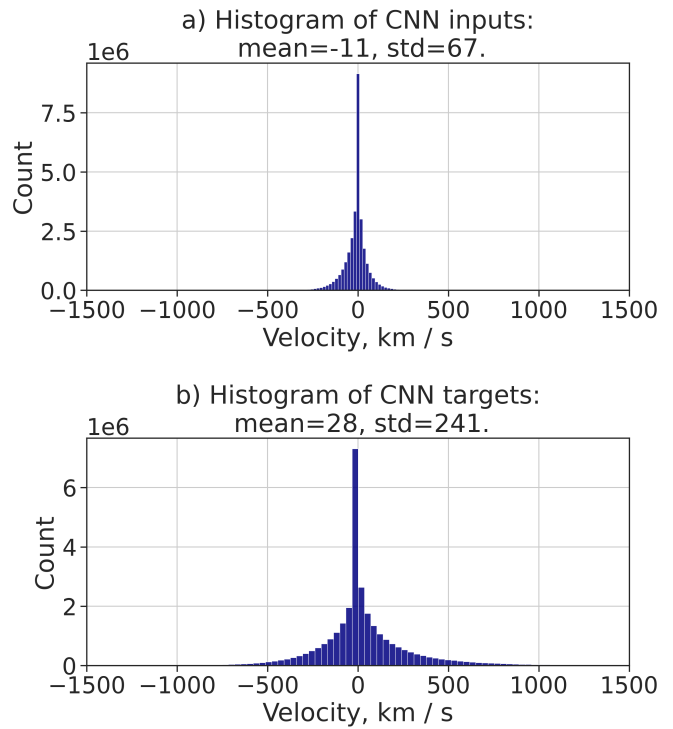
\plot{hists}{width=\columnwidth}{Histograms of velocity perturbations in the input and target datasets for training the CNN: (a) high-wavenumber model updates (10-th iteration, CNN inputs); (b) low-wavenumber model updates (CNN targets).}
Thus, considering the information on the CNN inputs and CNN outputs distribution, we rescale the dataset to have a zero mean and standard deviation of one. This is needed to enable a balanced training of the deep learning model.

\subsubsection{Network configuration}
In this study, we selected the FusionNet \citep{fusionNet} as an architecture of choice since it performed best in our preliminary experiments.
FusionNet is based on the architecture of a convolutional autoencoder as well as the U-net\citep{ronneberger2015u} architecture. One major difference between FusionNet and U-net architectures is the way in which skip connections are used. Namely, both architectures contain long skip connections, however FusionNet contains also short skip connections, contained in each residual block, which serve as a direct connection from the previous layer within the same encoding or decoding path. These nested short skip connections permit information flow across levels. Additionally, replacing concatenation with addition in long skip connections allow FusionNet to become a fully residual network, which resolves some common issues in deep neural networks (i.e., gradient vanishing).

The number of filters in each convolutional layer was increased to 110, which led to an architecture with $\sim$231~M of trainable parameters. Also, we added dropout layers (with a threshold value of 0.3) after each block of convolutional layers. The network optimization was driven by an Adam optimizer \citep{kingma2014adam}, with a learning rate of $0.0002$, and hyperparameters $\beta_1=0.5$ and  $\beta_2=0.999$.
Since the generation of a synthetic dataset requires multiple runs of FWI, we seek a way to expand the dataset by adding data augmentation. We horizontally (left-to-right) flipped input and target data to double the size of the training dataset. Altogether, this simple data augmentation and dropout layers in the network architecture reduced overfitting and improved the generalization capability of the network.
To limit the picture size for the CNN input layer, we resize input and target data to a size of $512 \times 128$ to simplify image propagation through the standard network blocks. 

\subsubsection{Metrics}
We use the $R^{2}$ score as a metric to compare predicted and target initial velocity models for evaluating performance of the inference and evaluating final FWI results. This metric measures how much of the variance is explained by the trained NN compared to the mean of the data. 
Assuming the true data - $y$ and the predicted data - $f$, the $R^{2}$ score formulates as follows:
\begin{align*}
R^2=1-\frac{SS_{res}}{SS_{tot}}, \\
SS_{res}=\sum_{i}{(y_i-f_i)^2}, \\
SS_{tot}=\sum_{i}{ (y_i-\overline{y})^2 },
\end{align*}
where $SS_{res}$ is the residual sum of squares, $SS_{tot}$ the total sum of squares. From the equations above, it follows that the $R^{2}$ score can vary from negative values to $+1$, where $1$ is the ideal match.
The $R^{2}$ value is comparable for different problems because it is normalized by the total sum of squares, which is proportional to the variance of the data.
\subsection{Training}

Training and testing of the neural network performance is a crucial step in deep learning and we explain our approach in more detail here.
In a subsequent section, we will then illustrate the quality of the CNN-predicted initial models by performing a conventional FWI with them. 

\subsubsection{Training CNN}

Our DL dataset was created by running 10 FWI iterations for each of the 3,904 randomly generated subsurface models. We split the full dataset, allocating 400 samples for the validation partition and 3,504 samples for the training partition. By horizontal-flip augmentation, we effectively double the amount of training and validation data. This led to the size of training and validation datasets of 7,008 and 800 samples, respectively. We use an $L^1$ training loss with a batch size of 4, and an early stopping callback to avoid overfitting.

To test the performance of the resulting neural network, we graphically show the CNN inference on the testing samples, i.e., the two benchmark velocity models and the field data. The results are presented for the benchmark models Marmousi II (\figref{/fwi_results/model__Marmousi_1d_lin_300_f_z}) and Overthrust (\figref{/fwi_results/model__Overthrust_1d_lin_300_f_z}), and for the marine streamer data (\figref{/fwi_results/model__cgg_lin_vp_long_300_f_z}).
Each figure represents a sequential application of the CNN to the presented velocity models, i.e. the inference result from the corresponding initial models.

We calculate the difference between presented initial velocity models and true velocity models with an $R^{2}$ score on the benchmark velocity models (Marmousi II, Overthrust).

\subsubsection{Testing CNN performance on Marmousi model}

The Marmousi II velocity model \citep{marmousi2} is an extended version in width and depth of the original Marmousi velocity model. It is based upon geology from the North Quenguela Trough in the Quanza Basin of Angola, Africa. 
The section is primarily composed of shale units, with occasional sand layers. The core of the complex faulted area is an anticline that is composed of marl. An unconformity and a partially evacuated salt layer separate the marls from the deeper anticlinal units, which are also mostly shales with some sand.
Generally, the Marmousi II model contains an anticline structure, interrupted with salt layers and faults. In the central part, there are hydrocarbon traps.

We do not expect the geology corresponding to the field data from north-western off-shore Australia to be so complicated as Marmousi II geology. Namely from the Australian seismic data and well log we do not see any signs of salt and anticline structures, generally we can conclude that Australian data represents relatively flat structure.
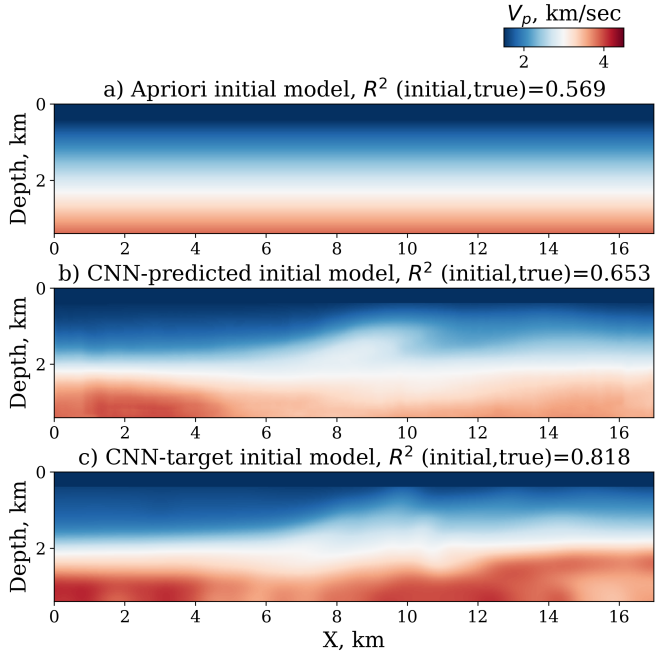
\plot{/fwi_results/model__Marmousi_1d_lin_300_f_z}{width=\columnwidth}{CNN inference on the Marmousi testing sample: (a) 1D initial model, (b) CNN inference result from the 1D initial model, (b) target initial model.}
In \figref{/fwi_results/model__Marmousi_1d_lin_300_f_z}, we present the process of prediction of the initial model for FWI. We compare the initial models with the true model using the $R^{2}$ score. 
\figref{/fwi_results/model__Marmousi_1d_lin_300_f_z}a represents the a priori initial model for FWI, the $R^{2}$ score accuracy is 0.569. \figref{/fwi_results/model__Marmousi_1d_lin_300_f_z}b represents the CNN inference result from \figref{/fwi_results/model__Marmousi_1d_lin_300_f_z}a. \figref{/fwi_results/model__Marmousi_1d_lin_300_f_z}c represents the CNN-target initial model.
The central anticline part of the true model appears in the predicted model. Furthermore, it does not contain any extra false features, in other words artifacts. The CNN-predicted initial model \figref{/fwi_results/model__Marmousi_1d_lin_300_f_z}b contains most of features of the CNN-target initial model, though the deep salt layers are not brightly expressed.
 
\subsubsection{Testing CNN performance on Overthrust model}
The Overthrust model depicts a complex thrusted stratigraphy, unconformably overlying an earlier extensional and rift sequence \citep{description_overthrust}. The three-dimensional Overthrust model contains three major zones: a monoclinal zone, a thrust anticline and a flat zone. The 2D section from the 3D Overthrust model that is used here contains a thrust anticline and a flat zone. This model is challenging for FWI, because of the low-velocity layers, squeezed between high-velocity ones. Our intention is to test the network in this geological setting even though it is not similar to the geology from Australian data.
In \figref{/fwi_results/model__Overthrust_1d_lin_300_f_z}, we show the estimation of the initial model for Overthrust. The CNN improved the $R^{2}$ score accuracy from 0.451 on the a priori 1D initial model to 0.497 on the final predicted initial model, while the CNN-target initial model $R^{2}$ score accuracy is 0.678. The CNN predicted a smooth anticline structure, while the a priori initial model was missing it. 
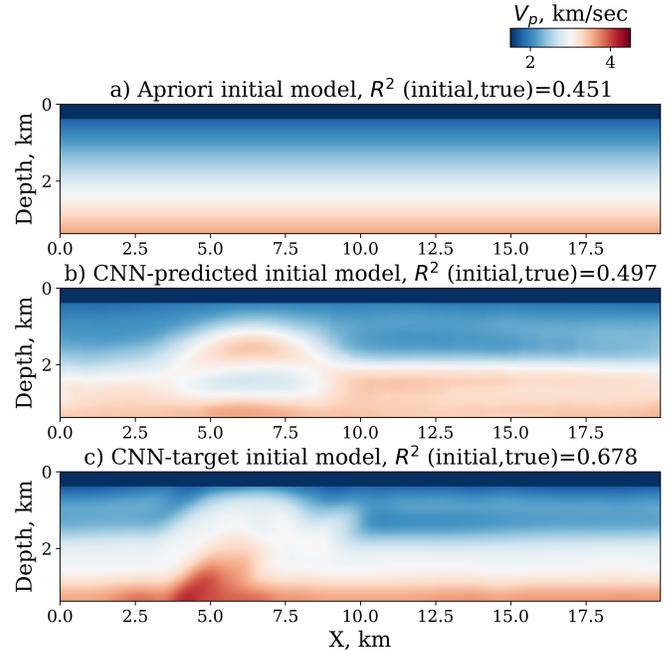
\plot{/fwi_results/model__Overthrust_1d_lin_300_f_z}{width=\columnwidth}{CNN inference on the Overthrust testing sample: (a) 1D initial model, (b) CNN inference result from the 1D initial model, (b) target initial model.}

\subsubsection{Testing CNN performance on field data}

Since we designed the training dataset to match parameters of the field data, we expect that the basic knowledge transfer between synthetic and field data becomes manageable for the CNN network. 
The inference of the CNN to field data is shown in \figref{/fwi_results/model__cgg_lin_vp_long_300_f_z}. The black dashed line on \figref{/fwi_results/model__cgg_lin_vp_long_300_f_z} represents the location of the wellhead of the well for which an acoustic $V_p$ log is available. The black solid line in \figref{/fwi_results/model__cgg_lin_vp_long_300_f_z} represents the $V_p$ acoustic well log itself.
The CNN inference introduces a complicated layered structure into the 1D linear trend of the a priori initial model. The predicted initial model contains high-velocity layers at depths 1.7 km and 3.5 km. The layer at depth 1.7 km correlates with the $V_p$ acoustic log.

The verification of the predicted initial model with only single borehole log is hardly enough to understand whether the initial model is appropriate for FWI, but it is always important to use all available information for the field data processing.

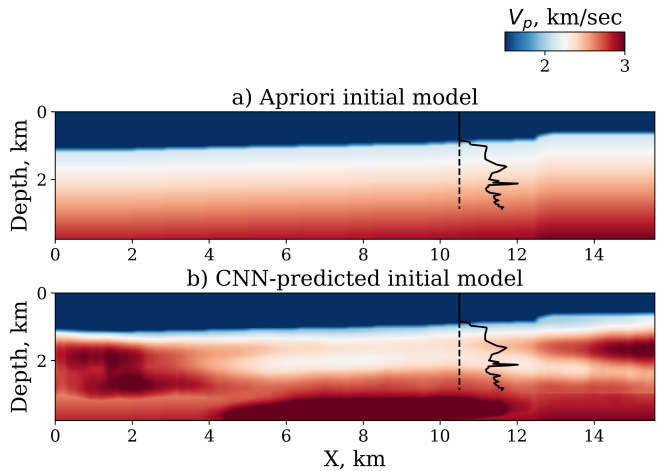
\plot{/fwi_results/model__cgg_lin_vp_long_300_f_z}{width=\linewidth}{CNN inference on field data testing sample: (a) 1D initial model, (b) CNN inference result from the 1D initial model. Dashed line - location of the wellhead, solid line - $V_p$ acoustic well log.}
\section{FWI results}

In this section, we are interested to see if the inferred initial model predictions from the DL framework improve the FWI results on seismic data with missing low frequencies. Particularly, we investigate if the CNN-predicted initial model leads to better inversion results than starting from a 1D initial model, and how it compares in the benchmark cases to the ideal case where FWI uses a smoothed true model as an initial model.

We developed a special inversion workflow for performing numerical experiments from different initial models on the data with lack of low frequencies.
First, different from the approach for generation of CNN-input data we chose global correlation norm misfit \citep{choi2012} as more robust, than $L_2$ misfit. Then, we divide iterative inversion process to five stages of increasing low-pass filter cut-off frequencies in the following order: 6 Hz, 6 Hz, 7 Hz, 8 Hz, 8 Hz.
At the same time, in each stage we apply spatial filtering of gradients similar to \citep{ravaut2004multiscale} changing damping parameter $WD_{damp}$ in following order: 1.5, 1, 0.1, 0.25, 0.125. Higher value of this parameter means stronger damping and smoother gradients.

We will use the $R^2$ score as a similarity measure for the velocity models. For each of the following figures, we present the similarity scores between the displayed velocity model and the true model in the titles. In the following figures initial models for FWI are shown in the left column, the results of FWI are shown in the center column, and in the right column the corresponding FWI data misfit functions in the same absolute amplitude scale are displayed. The first row shows the results corresponding to the a priori 1D initial model, the second row corresponds to the FWI on CNN-predicted initial model, and the third row corresponds to the FWI on CNN-target initial model.


\subsubsection{Marmousi}

\figref{fwi_results/Marmousi_final} shows the inversion results from high-frequency data for the Marmousi benchmark model.
FWI started from the 1D a priori initial model decreased the $R^2$ score from an accuracy of 0.569 to a final 0.508, \figref{fwi_results/Marmousi_final}a and \figref{fwi_results/Marmousi_final}b, respectively.
FWI started from the CNN-predicted initial model improved the accuracy from 0.653 to a final 0.694, \figref{fwi_results/Marmousi_final}d and \figref{fwi_results/Marmousi_final}e, respectively.
FWI started from the CNN-target initial model improved the accuracy from 0.818 to 0.870, \figref{fwi_results/Marmousi_final}g and \figref{fwi_results/Marmousi_final}h, respectively.

FWI with the CNN-target initial model produces obviously the best result which can be used as an ideal reference for other initial models. The image shows no cycle-skipping artifacts (\figref{fwi_results/Marmousi_final}h) and the $R^2$ score accuracy 0.870 is close to the ideal value of $+1$.
The CNN-predicted initial model starts with an $R^2$ score accuracy of 0.653, confirming that its central anticline part is well predicted. In the resulting final FWI image, the central part shows a comparable result as in the ideal case, but still misses accuracy in the deeper and boundary areas.
The worst inversion result was obtained from the a priori initial model, with a final $R^2$ score accuracy of 0.508. The inverted model \figrefp{fwi_results/Marmousi_final}b contains cycle-skipping artifacts, highlighting the importance of a good starting model for such a demanding subsurface structure.
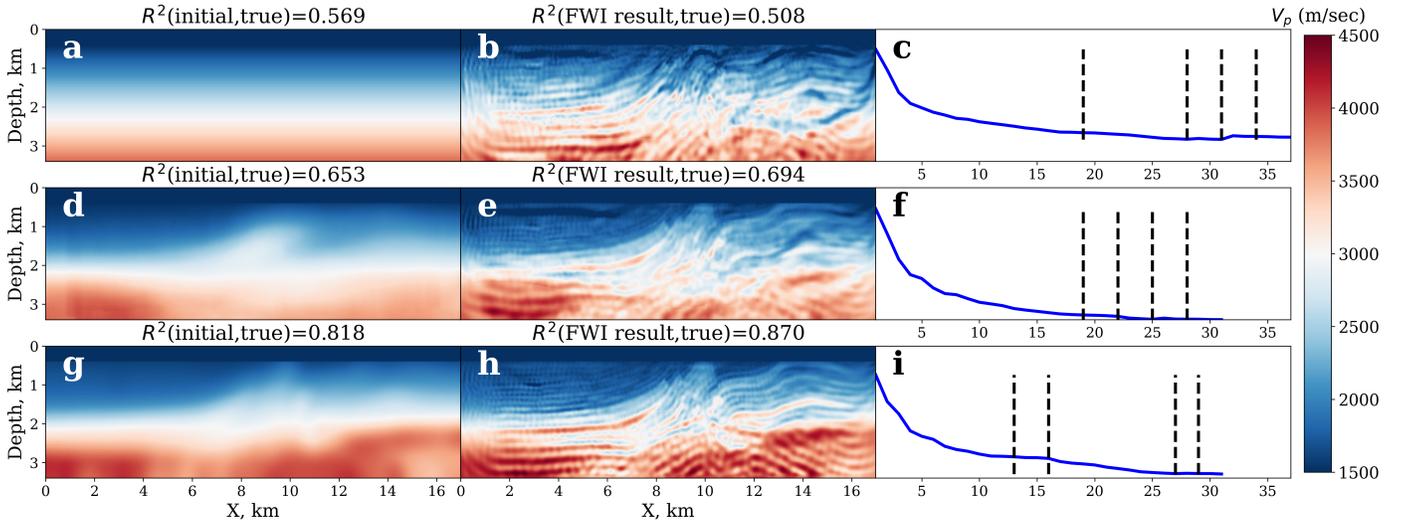
\fullplot{fwi_results/Marmousi_final}{width=\linewidth}{High-frequency band FWI on Marmousi data. Left column - initial models for FWI, center column - FWI-inverted models, right column - FWI data misfit functions. a) 1D initial model, d) CNN-predicted initial model, g) target initial model (smoothed true). Black dashed lines on the plots of FWI data misfit functions c), f), i) represent boundaries between different FWI stages.}
\subsubsection{Overthrust}

The FWI results for the Overthrust benchmark model are shown in \figref{fwi_results/Overthrust_final}.
FWI started from the 1D initial model increased the $R^2$ score from an initial value of 0.451 to a final 0.458 (\figref{fwi_results/Overthrust_final}a and \figref{fwi_results/Overthrust_final}b respectively).
FWI started from the CNN-predicted initial model increased the $R^2$ score from 0.497 to 0.526, \figref{fwi_results/Overthrust_final}d and \figref{fwi_results/Overthrust_final}e, respectively.
FWI started from the CNN-target initial model decreased the $R^2$ score from 0.678 to 0.674, \figref{fwi_results/Overthrust_final}g and \figref{fwi_results/Overthrust_final}h.

FWI with the CNN-target initial model can be seen as an ideal reference case, which produces the best result like in the Marmousi model test.
Still, the inverted result (\figref{fwi_results/Overthrust_final}h) is far from being perfect as it contains many cycle-skipping artifacts, especially in the deeper part of the model. This demonstrates that the Overthrust benchmark is even more challenging for FWI to obtain reliable results from high-frequency data than the Marmousi test case.

The CNN-predicted initial model starts with an $R^2$ score of 0.497, which is smaller than the $R^2$ score for the CNN-target initial model. The error in inference occurs mainly in the central part of the model containing a smooth thrust anticline. Though the inference is not perfect, the prediction accuracy still allows for a small improvement of the FWI result accuracy to a final $R^2$ value of 0.526. However, the final image contains various inversion artifacts and hardly reaches the ideal case image. This demonstrates that especially the anticline structure is an essential part of the initial model estimation, and strongly required in the absence of low-frequency data to constrain FWI results.
The worst inversion result has been obtained from the a priori initial model and reached the $R^2$ score of 0.458. The inverted model contains many cycle-skipping artifacts, missing most of the anticline structures. Thus, the most important part of the Overthrust model, i.e., the thrust anticline, is not inverted at all.

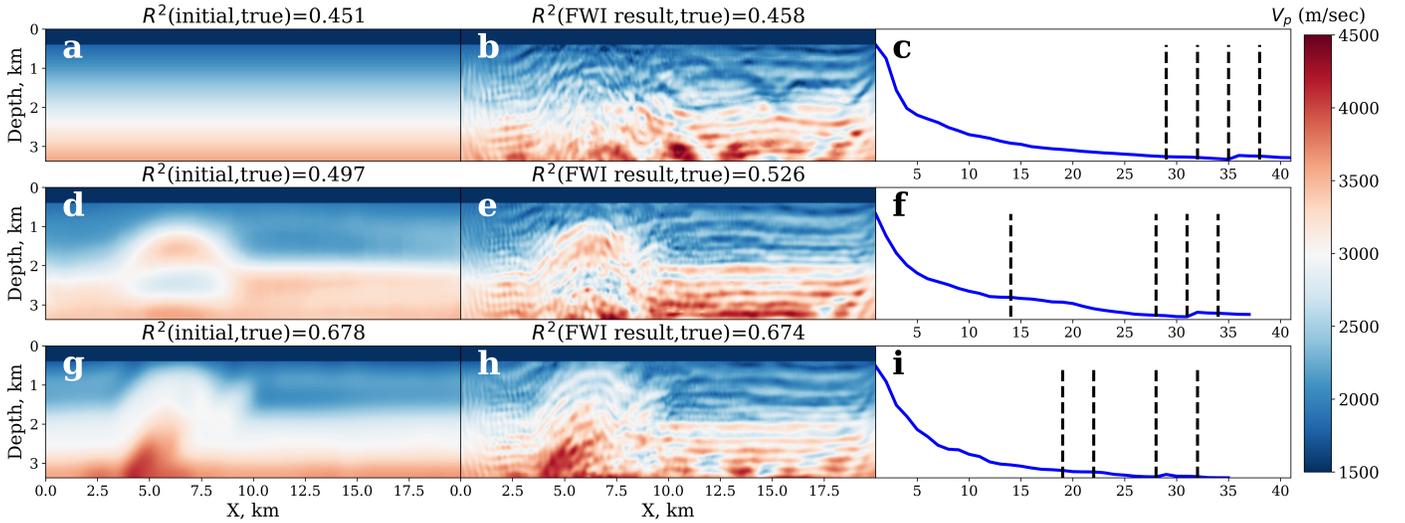
\fullplot{fwi_results/Overthrust_final}{width=\linewidth}{High-frequency band FWI on Overthrust data. Left column - initial models for FWI, center column - FWI-inverted models, right column - FWI data misfit functions. a) 1D initial model, d) CNN-predicted initial model, g) target initial model (smoothed true). Black dashed lines on the plots of FWI data misfit functions c), f), i) represent boundaries between different FWI stages.}

For both benchmark models, Marmousi and Overthrust, we observe that the data misfit for the FWI started from CNN-predicted initial models decreases to a smaller value than the FWI started from the a priori initial models. Obviously, the misfit for the ideal result where FWI was started from the true smooth initial model decreases to the smallest value.
We find that according to the $R^2$ score, the CNN-predicted models are closer to the true model than the initial a priori 1D models. This is already an encouraging outcome, highlighting that the CNN inference is able to predict some meaningful initial models. Given that both benchmark models, Marmousi and Overthrust, have not been included in the training and according to \figref{random_v} might even contain velocity logs outside of the training velocity range, the FusionNet neural network seems to be able to generalize and find a corresponding smooth low-wavenumber model based on its high-wavenumber input data. Also, the FWI results for the CNN-predicted initial models produce consistently better FWI results than for the a priori 1D ones.

\subsubsection{Field data}
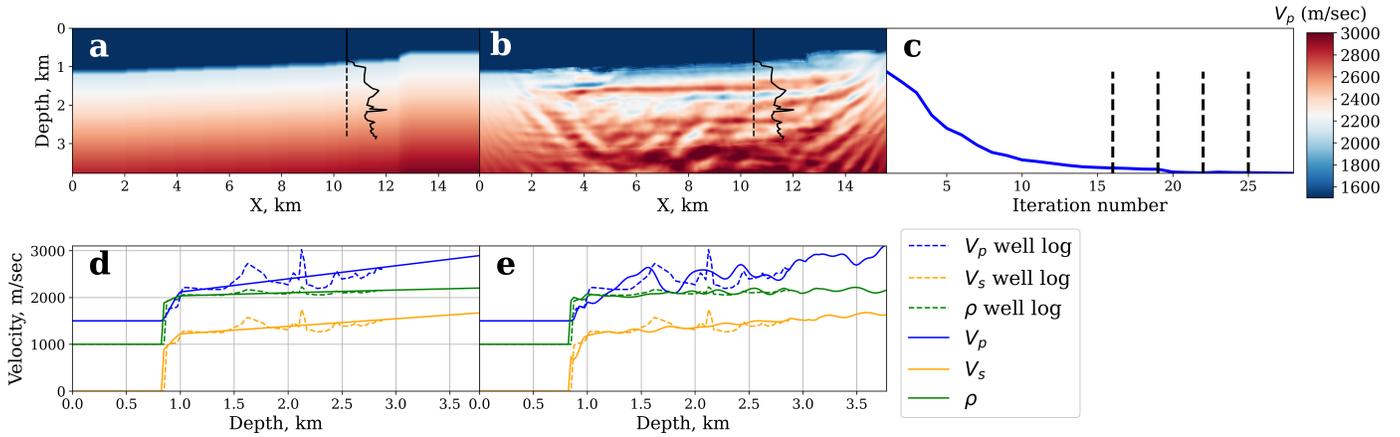
\fullplot{fwi_results/cgg_1d_highfreq}{width=\linewidth}{High-frequency band FWI on marine streamer data from the 1D initial model. a) 1D initial model for FWI, b) FWI-inverted model, c) FWI data misfit function, d) comparison of initial model log vs well log from 10.5~km of field survey, e) comparison of FWI-inverted log vs well log from 10.5~km of field survey.}
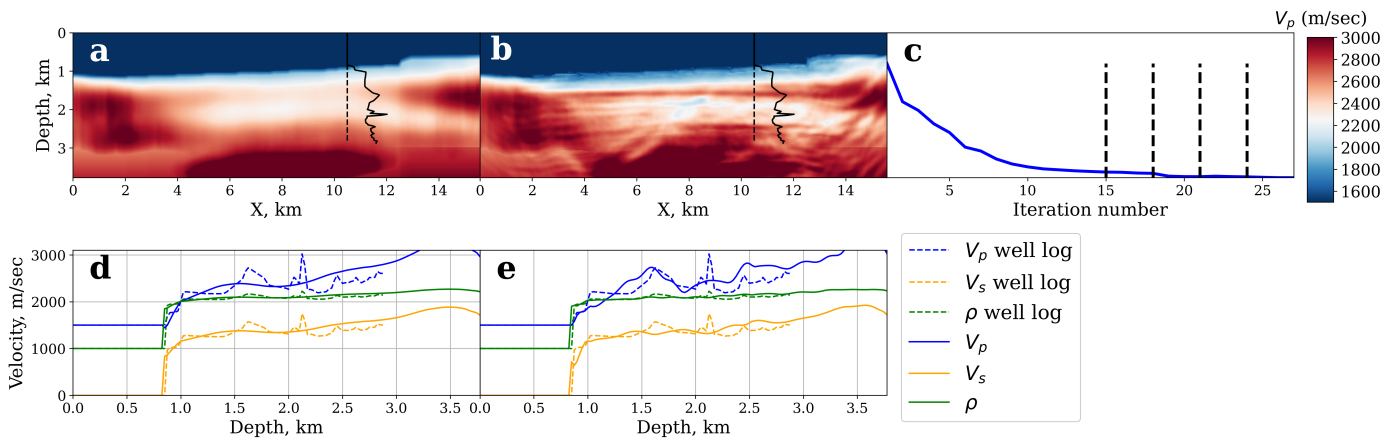
\fullplot{fwi_results/cgg_cnnpred_highfreq}{width=\linewidth}{High-frequency band FWI on marine streamer data from the CNN-predicted initial model. a) CNN-predicted initial model for FWI, b) FWI-inverted model, c) FWI data misfit function, d) comparison of initial model log vs well log from 10.5~km of field survey, e) comparison of FWI-inverted log vs well log from 10.5~km of field survey.}

To validate our approach on the field data acquired by CGG in the northwestern part of Australian shelf, we interpret the FWI result using an acoustic $V_p$ well log from a 10.5~km offset of the field survey, denoted as a black solid line in \figref{fwi_results/cgg_1d_highfreq} and \figref{fwi_results/cgg_cnnpred_highfreq}. Unfortunately, this is the only well log available for this particular dataset. We recalculate acoustic $V_p$ well log with Gardner relation to obtain the $V_s$ and $\rho$ well logs. In \figref{fwi_results/cgg_1d_highfreq} and \figref{fwi_results/cgg_cnnpred_highfreq} $V_p$, $V_s$, $\rho$ are shown with blue, yellow and green colors, respectively; well logs are shown with a dashed line, inverted logs are shown with solid lines.

To illustrate the effectiveness of our DL method, we first run FWI on seismic data with missing low frequencies starting from the a priori initial model and the CNN-predicted initial model. The results are presented in \figref{fwi_results/cgg_1d_highfreq} for the a priori 1D initial model and \figref{fwi_results/cgg_cnnpred_highfreq} for the CNN-predicted initial model, respectively.

In \figref{fwi_results/cgg_1d_highfreq}b, we observe that FWI result contains cycle-skipping artifacts. On the $V_p$ log comparison in the \figref{fwi_results/cgg_1d_highfreq}e, we observe that the inverted log correlates with the well log poorly.

In \figref{fwi_results/cgg_cnnpred_highfreq}a and \figref{fwi_results/cgg_cnnpred_highfreq}d, we observe that the CNN predicted smooth high-velocity anomalies at depths 1.7 km and 3.5 km.
Generally, the inversion result from the CNN-predicted initial model seems to have fewer cycle-skipping artifacts than the inversion result from the a priori initial model. Inversion from the CNN-predicted initial model restored thinner geologic layers than the a priori initial model.
The FWI result from the CNN-predicted initial model \figref{fwi_results/cgg_cnnpred_highfreq}b,e becomes better than the result from the a priori 1D initial model \figref{fwi_results/cgg_1d_highfreq}b,e. High-velocity layers from the well log at depths 1.7 km and 2.2 km are inverted better using the CNN-predicted initial model. 
However inversion results from the CNN-predicted initial model are still not perfect. The layer at 1.1 km is not inverted probably because of cycle-skipping, the same can be concluded for the layer on depth 2.2 km. Also velocities below 2.2 km are overshooted, because of the too high velocity anomaly introduced by CNN on the depths below 3 km.



\section{Discussion}
The substantial importance for this approach is the a priori 1D initial model, for which we choose to generate the CNN data and then predict low wavenumbers. For inaccurate 1D a priori initial models, the CNN provides a significant increase in the accuracy. For the accurate 1D a priori initial models, which are very close to the true smoothed model, the CNN does not result in considerable increase in the accuracy.

The tuning of FWI workflows for synthetic and field data that lack of low frequencies took a lot of time and effort because it is challenging to run multiscale FWI starting from 5~Hz. Nevertheless, for the CNN-predicted initial model we consider the results of multiscale FWI starting from 5~Hz not satisfactory because the CNN does not predict the initial model accurately enough to avoid cycle-skipping.

The approach for increasing the number of filters and dropout in FusionNet proved to be working in terms of accuracy. 
The standardization scaling proved to be working better than min-max scaling.
\section{Conclusions}

In this study, we looked from a different point of view at the problem of building the initial low-wavenumber model for full-waveform inversion by deep learning. We constructed a mapping between the high-wavenumber FWI model updates and the initial model for FWI by a deep learning neural network to deal with the lack of a priori knowledge and low-frequency seismic data information. In numerical experiments on the benchmark models and field dataset the CNN-predicted initial model led to better FWI results than the 1D initial model.
Our method can be seen as an alternative to initial velocity model building or it can complement existing approaches, such as FWI model regularization techniques, i.e., gradient filtering and conditioning, low-frequency seismic data extrapolation, or ray-based seismic tomography. 
The proposed method relies on full-waveform inversion for constructing the input for CNN inference. Therefore, the technique requires a significant amount of computational resources to generate a training dataset. However, once the network is trained on a dataset, the inference on new models, with similar acquisition parameters and FWI implementation, can be produced fast without retraining. In this study we demonstrated the feasibility of the proposed DL approach, however, further research is still required to evaluate and improve our method for exploration applications.

\section{Acknowledgement}
We thank members of the Seismic Modelling and Inversion group (SMI) and the Seismic Wave Analysis Group (SWAG) at KAUST for constructive discussions on deep learning and modeling. In addition, we are grateful to the members of the KAUST KSL team for valuable discussions and technical support. 
Computational resources were provided by KAUST's supercomputing facility Shaheen II.
\newline
We acknowledge CGGVeritas Services SA for providing confidential field marine data from the North-Western Australian offshore.
\newline
Open-access data to reproduce the paper will be available on Github \url{https://github.com/pplotn/dl_fused_fwi.git}. The field data can not be made publicly available.
 
\newpage
\bibliography{main}
\bibliographystyle{seg}  
\end{document}